\documentclass{article} 
\usepackage{iclr2021_conference,times}

\usepackage{amsmath,amsfonts,bm}

\def\eqref#1{equation~\ref{#1}}

\def\1{\bm{1}}

\DeclareMathAlphabet{\mathsfit}{\encodingdefault}{\sfdefault}{m}{sl}
\SetMathAlphabet{\mathsfit}{bold}{\encodingdefault}{\sfdefault}{bx}{n}

\usepackage{hyperref}
\usepackage{url}
\usepackage{graphicx}

\title{Foundation AI Model for Medical Image Segmentation}

\usepackage{authblk}

\author[1,2]{Rina Bao}
\author[1,2]{Erfan Darzi}
\author[1,2]{Sheng He}
\author[1,2]{Chuan-Heng Hsiao}
\author[1,2]{Mohammad Arafat Hussain}
\author[1,3]{Jingpeng Li}
\author[3]{Atle Bj\o{}rnerud}
\author[1,2]{Ellen Grant}
\author[1,2]{Yangming Ou}

\affil[1]{Boston Children's Hospital}
\affil[2]{Harvard Medical School}
\affil[3]{University of Oslo}

\iclrfinalcopy 
\begin{document}

\maketitle

Foundation models refer to artificial intelligence (AI) models that are trained on massive amounts of data and demonstrate broad generalizability across various tasks with high accuracy. These models offer versatile, one-for-many or one-for-all solutions, eliminating the need for developing task-specific AI models. Examples of such foundation models include the Chat Generative Pretrained Transformer (ChatGPT) and the Segment Anything Model~\citep{kirillov2023segment} (SAM). These models have been trained on millions to billions of samples and have shown wide-ranging and accurate applications in numerous tasks such as text processing (using ChatGPT) and natural image segmentation (using SAM).

\begin{figure*}[htb]
    \centering
    \includegraphics[width=\textwidth]{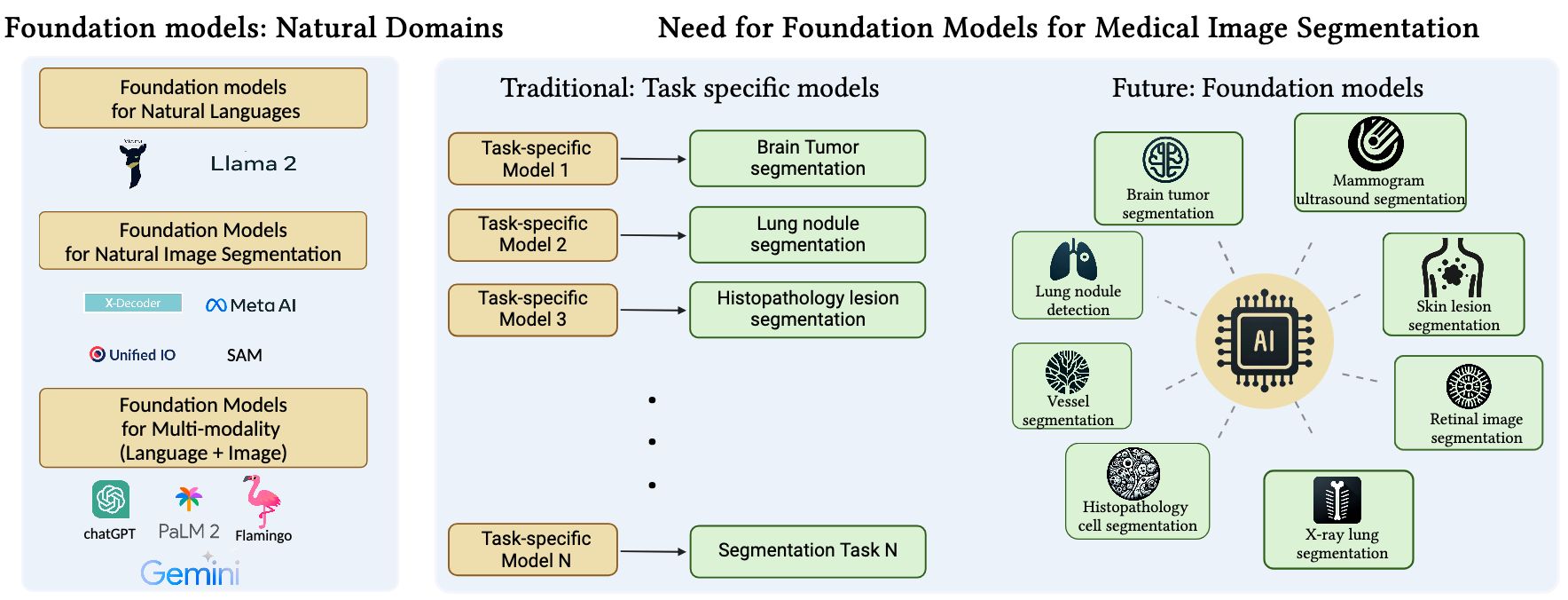}
    \caption{The need for foundation models in the medical image segmentation domain. The left panel lists foundation models in natural language processing, natural image segmentation, or multi-modal (language, image, etc.) tasks. The right panel shows the need to transition from tens of thousands of task-specific AI models to a foundation for medical image segmentation tasks.}
    \label{fig:need}
\end{figure*}

\begin{figure*}[htb]
    \centering
    \includegraphics[width=0.9\textwidth]{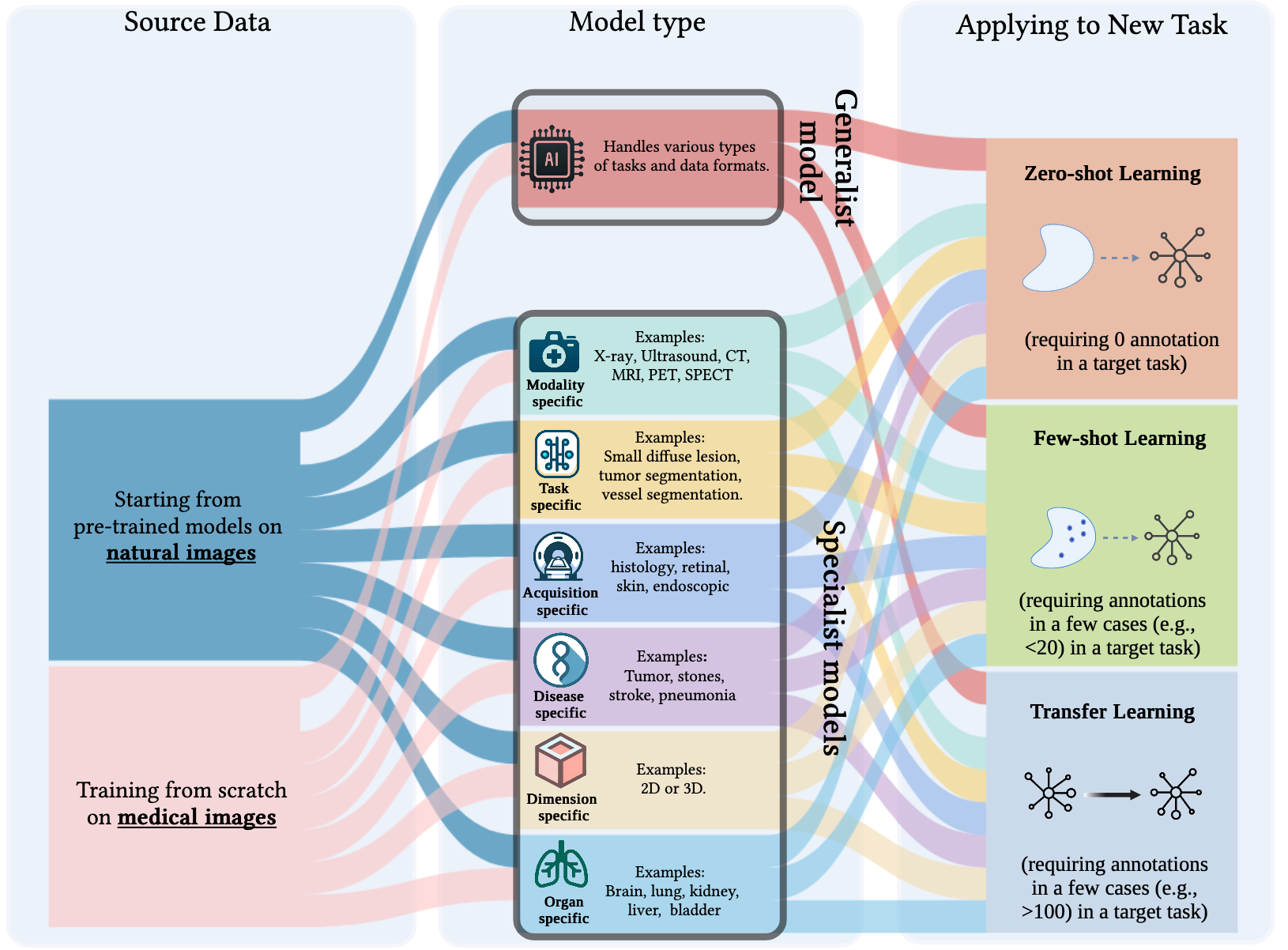}
    \caption{Paths for creating and applying foundation models in medical image segmentation.}
    \label{fig:overview}
\end{figure*}

In medical image segmentation – finding target regions in medical images – there is a growing need for these one-for-many or one-for-all foundation models (as illustrated in Figure~\ref{fig:need}). Such models could obviate the need to develop thousands of task-specific AI models, which is currently standard practice in the field. They can also be adapted to tasks with datasets too small for effective training. We discuss two paths to achieve foundation models for medical image segmentation and comment on progress, challenges, and opportunities. One path is to adapt or fine-tune existing models, originally developed for natural images, for use with medical images. The second path entails building models from scratch, exclusively training on medical images (as illustrated in Figure~\ref{fig:overview}).

\section*{Source Data--Natural or Medical Images?}

This first path (Figure~\ref{fig:overview}, top box in the first column) takes advantage of the facts that a) taking a natural image, often by cellphones or cameras, is less costly than acquiring a medical image, often by special medical devices; b) many datasets of natural images with ground-truth segmentations are publicly available, in larger amount than publicly-available medical images; and c) foundation models recently emerged for natural images with high generalizability and accuracy.

SAM is such a foundation model for natural image segmentation. It debuted in April 2023 and was trained on 11 million natural images with $>1$ billion ground-truth segmentation masks. It can segment a whole 2D image into an automatically- determined number of regions (SAM-Semantic). It can also take prompts of seeding points (SAM- Point) or bounding boxes (SAM-Box) for user- interactive segmentation. In 26 datasets in the original paper and many more data in other independent tests, SAM, even without seeing any data from a new natural image dataset (the so-called zero-shot application), has achieved as high, if not higher, accuracies compared to AI models that have seen part of the data in the new natural image dataset (the so-called task-specific AI models).

The promising generalizability and accuracy in natural images provoke applications of SAM to medical images, mainly in three directions.
(1) Directly applying SAM to medical images: contrary to its performance in natural images, SAM shows Dice coefficients as much as 0.5-0.7 lower than those from task-specific models – in pathology images~\citep{deng2023segment}, liver tumor segmentation~\citep{hu2023sam}, brain MRI segmentation~\citep{bao2023boston}, abdominal CT organ segmentation~\citep{roy2023sam}, and numerous other medical image datasets~\citep{he2023computervision}. Major challenges occur in small target regions, low-contrast, irregular shapes, 3D images, and more on MRI/CT or other non-camera images. These difficulties arise from the unique texture and contrast characteristics of medical images, which differ significantly from those in natural images. (2) Adapting SAM to medical images: let SAM ``see" more medical images. Given SAM's large parameter size, researchers have concentrated on retraining smaller and often the last segments of SAM model while maintaining most of the original weights intact. Fine-tuning efforts have targeted specific applications like segmentation of skin cancer~\citep{hu2023skinsam}, polyp~\citep{zhou2023can}, and multi-organs~\citep{cheng2023sam}. The adaptions to specific medical images may, however, reduce the generalizability among other medical image segmentation tasks. (3) Automate SAM's prompts for medical image segmentation: approaches such as DeSAM~\citep{gao2023desam}, AutoSAM~\citep{shaharabany2023autosam}, All-in-SAM~\citep{cui2023all}, and others, aim to standardize prompts across medical tasks, reducing dependence on human input and improving stability in prompts.

A second source of annotated data could be solely medical images (Figure~\ref{fig:overview}, lower box in the first column) – Medical images differ significantly from natural images in aspects such as acquisition, organs involved, imaging modality, imaging principles, resolution, contrast, and dimension. Additionally, the segmentation targets in medical images may be small (e.g., $< 1\%$ of the entire image)~\citep{bao2023boston}, diffuse (multi-focal)~\citep{bao2023boston}, subtle (showing low contrast from neighboring structures), and heterogeneous.

\section*{Types of Foundation Models--Generalist or Specialist?}
One can pool vast medical image segmentation datasets with ground truth and use all these data to train a unified foundation medical model. We call them generalist models because all source images are used, regardless of imaging modality, dimension, acquisition, contrast, scanning organ, or the segmentation tasks (top box, second column in Figure~\ref{fig:overview}). If such ambitious generalist foundation models show unsatisfactory accuracies, a compromise is to build separate foundation models for a fraction of the medical images that share similar properties, either by organ, modality, task, dimensionality, or imaging principles. We can call such models specialist foundation models (Figure~\ref{fig:overview}, bottom panel in the second column). Possible specialist models can be: (1) Foundation model of the same organ. This approach focuses on a single organ, such as the brain or heart, but across various diseases or imaging modalities. For instance, a model might be trained to segment the liver or liver tumor in either abdominal CT or MRI images. (2) Foundation model of the same medical imaging modality. We can train one foundation model for all X-rays, and one foundation model for all ultrasound images, across different diseases or organs. The key here is to develop a model that understands the specific characteristics of the imaging modality (X-ray here for example) while being adaptable to the diverse anatomical structures and pathologies it may encounter. (3) Foundation model of similar tasks across organs, diseases, or modalities. An example would be segmenting small, diffuse diseases such as hypoxic-ischemic encephalopathy in infants, ischemic stroke lesions in adults, and multiple sclerosis. Despite the diseases being different, these brain lesion types share similar characteristics in medical images. (4) Foundation model for a certain image dimension. One may group all 2D images (e.g., X-ray, histology, retinal image, skin photos, endoscopic images, mammography, ultrasound, etc.) and build a 2D medical image segmentation foundation model. Similarly, we can merge all 3D images (e.g., MRI, CT, PET, SPECT, tomosynthesis, etc.) to build a 3D medical image segmentation foundation model. (5) Foundation model for certain acquisition sources: We can group medical images taken from cameras (e.g., histology, retinal, skin, endoscopic images, etc.) because they are closer to natural images in imaging principles. Similarly, we can build foundation models for medical images taken by radiation (e.g., X-ray, CT, PET, SPECT, and tomosynthesis), and foundation models for non-camera and radiation-free medical images (e.g., MRI, ultrasound, near-infrared spectroscopy, etc.).

\section*{Applying Foundation Models}
Foundation models, regardless whether starting from natural or medical images, regardless of whether being a generalist or specialist models, can be applied to a given medical image segmentation task in three ways, as shown in the right column in Figure~\ref{fig:overview}. The accuracies of each path can be tested in three settings: (1) zero-shot (most ambitious for the highest generalizability): applying the foundation model directly to an arbitrary medical image segmentation task, requiring no annotated data from this task~\citep{he2023computervision}; (2) few-shot: allowing the foundation model to see some (typically $<20$ subjects) examples in the new task~\citep{pachetti2023systematic}; (3) fine-tuning: feeding ample (usually \textgreater 100 subjects) examples with ground-truth annotations~\citep{zhang2023challenges}, gaining accuracy in this target task while demanding data for each new medical image segmentation task (i.e., reduced generalizability). As mentioned in the “Source Data” section, the three application scenarios have been reflected in applying SAM to new medical image segmentation tasks.

\section*{Challenges and Opportunities}
Foundation models in medical imaging inherit general challenges that are prevalent in machine learning. These challenges include, but are not limited to, the validity and intra-/inter-rater consistency of ground truth, AI model interpretation, data privacy, and practical deployment. In this context, we focus specifically on challenges unique to foundation models for medical image segmentation, rather than the challenges also shared by other task-specific AI models.

\textit{Sample size.} What is the sufficient sample size that can comprehensively cover the diversity of medical images? It took SAM over 1 million images to achieve high generalizability and accuracy, and that was just for 2D natural images. Given the unique challenges in medical images, a foundation model for medical image segmentation may need more samples than for natural image segmentation. Despite challenges, opportunities exist as well. Today, hundreds of thousands of medical images with ground-truth masks are publicly available~\citep{cheng2023sam2d}, allowing us, at least to some extent, to test the effect of sample sizes.

\textit{Data preparation.} Merging medical image datasets is a non-trivial task. Trained exclusively on 2D natural images, SAM normalized all natural images to 1024×1024 pixels, 3-channel colors (Red, Green, and Blue), and 0-255 intensity in each channel. However, medical images are more complicated. They come from different acquisitions – cameras (retina, histology, endoscopy, etc.), radiations (X-ray, CT, PET, etc.), ultrasound, and magnetic resonance (MRI). They vary in dimension (2D, 3D), the number of channels (e.g., single- or multiple MRI sequences), spatial resolution (micrometers in histology versus millimeters in MRI), and many other factors. Therefore, merging medical images requires standardization of resolution, dimension, intensity, and color in medical images from different datasets. How the merge and pre-processing impact the final accuracy and generalizability remains to be studied.

\textit{Prompt Design.} Segmenting natural images is for general users while segmenting medical images is more often for healthcare professionals. This raises pivotal questions: Is the need for user prompts essential, or can full automation effectively address the task? If prompts are necessary, what forms should they take? Possibilities include points, contours, boxes, arrows, clinical patient records, and text chats, among others. Health professionals may also request interactive prompts to modify the auto-processed results. In terms of application, discerning which tasks necessitate generalist models versus specialist models, as well as identifying tasks best suited to zero-shot, few- shot, or transfer learning approaches, becomes paramount. This raises the question of whether a case-by-case ad hoc testing approach is sufficient. Or ideally, there should be established guidelines for ``learning to learn".

In summary, the recent debut and success of foundation models in natural image segmentation further fuel the desire to develop foundation models for medical image segmentation tasks. At least three major modules need to be studied: source data (natural or medical), type of foundation models (generalist or specialist), and application (zero-shot, few-shot, or fine-tuning). With the growing public availability of medical data, the standardization of data preparation, and basic prompt designs of segmentation, this direction may soon see an explosion of studies.

\bibliography{sn-bibliography}
\bibliographystyle{iclr2021_conference}

\appendix

\end{document}